\documentclass[english,twocolumn,prl,aps,groupedaddress,showpacs]{revtex4-1}
\usepackage{lmodern}
\usepackage[T1]{fontenc}
\usepackage[latin9]{inputenc}
\usepackage{babel}
\usepackage{amsmath}
\usepackage{amssymb}
\usepackage{graphicx}
\usepackage{esint}
\usepackage[unicode=true,
 bookmarks=true,bookmarksnumbered=false,bookmarksopen=false,
 breaklinks=false,pdfborder={0 0 1},backref=false,colorlinks=false]
 {hyperref}
\hypersetup{pdftitle={Theory of unitary Bose gases},
 pdfauthor={J.J.R.M. van Heugten and H.T.C. Stoof},
 hidelinks}
\usepackage{bbold}

\begin{document}

\title{Theory of unitary Bose gases}

\author{J.J.R.M. van Heugten and H.T.C. Stoof}

\affiliation{Institute for Theoretical Physics, Utrecht University, Leuvenlaan
4, 3584 CE Utrecht, The Netherlands}
\begin{abstract}
We develop an analytical approach for the description of an atomic
Bose gas at unitarity. By focusing in first instance on the evaluation
of the single-particle density matrix, we derive several universal
properties of the unitary Bose gas, such as the chemical potential,
the contact, the speed of sound, the condensate density and the effective
interatomic interaction. The theory is also generalized to describe
Bose gases with a finite scattering length and then reduces to the
Bogoliubov theory in the weak-coupling limit.
\end{abstract}


\maketitle
\textit{Introduction.} --- Although strongly interacting systems are
known to be notoriously difficult to describe from first principles,
remarkable progress in our understanding has been made in recent years.
For example, the universal nature of fermionic many-body systems with
resonant two-body interactions has been successfully studied experimentally
\citep{Bloch2008Manybody}. This has been achieved by utilizing the
high degree of control available in ultracold atomic gas experiments,
which allow the investigation of many-body systems from the weakly
to the strongly interacting case, by using a magnetic-field-tunable
Feshbach resonance \citep{Chin2010Feshbach,Bloch2008Manybody,Stoof2009Ultracold}.
The remarkable property of such resonant systems, which have an infinite
scattering length and are therefore said to be at unitarity, is that
at zero temperature there is no other length scale than the average
interatomic distance that is set by the particle density $n$. As
a result all thermodynamic quantities, when appropriately scaled,
can then be expressed in terms of a set of universal numbers. One
of the most crucial quantities of the Fermi gas at unitarity is the
chemical potential
\begin{equation}
\mu=(1+\beta)\epsilon_{F},\label{eq:Chemical_potential}
\end{equation}
which is given by an universal constant times the Fermi energy $\epsilon_{F}=\hbar^{2}k_{F}^{2}/2m$,
where $k_{F}=(6\pi^{2}n/2s+1)^{1/3}$ is the Fermi momentum and $s=1/2$
due to the hyperfine degrees of freedom. The universal constant $\beta$
can be interpreted as describing the deviation from the ideal gas
result due to interactions and was found to be $\beta\simeq-0.63$
experimentally as well as theoretically \citep{Ku2012Revealing,Zurn2012Precise}.

Not only fermionic many-body systems are studied with ultracold atomic
gases, but also its bosonic counterpart can be experimentally realized.
In recent years, there has therefore been an increasing interest in
the experimental study of the strongly interacting Bose gas at low
temperatures. It is expected on dimensional grounds that the Bose
gas at unitarity, if stable, has similar universal properties as that
of the unitary Fermi gas. For instance Eq.~(\ref{eq:Chemical_potential})
is expected to hold also but with $s=0$ and a different value of
$\beta$ due to the different statistics of the atoms. In contrast
to fermionic cold atomic gases, the experimental study of bosons with
increasing interaction strength is complicated by the loss of atoms,
as a consequence of a strong increase in the rate of inelastic three-body
recombination processes caused by the absence of the Pauli principle
and the existence of Efimov trimers. As a result, up to now only a
lower bound of $\beta>-0.56$ was determined experimentally \citep{Navon2011Dynamics}.
These inelastic three-body processes result in the formation of molecules,
which shows that the actual ground state of these gases is a Bose-Einstein
condensate of molecules. Nevertheless, it may still be experimentally
possible to create the meta-stable state of a Bose-Einstein condensate
of atoms at large scattering lengths for a sufficiently long time.
\begin{figure}[t]
\begin{centering}
\includegraphics[bb=10bp 10bp 375bp 226bp,clip,scale=0.68]{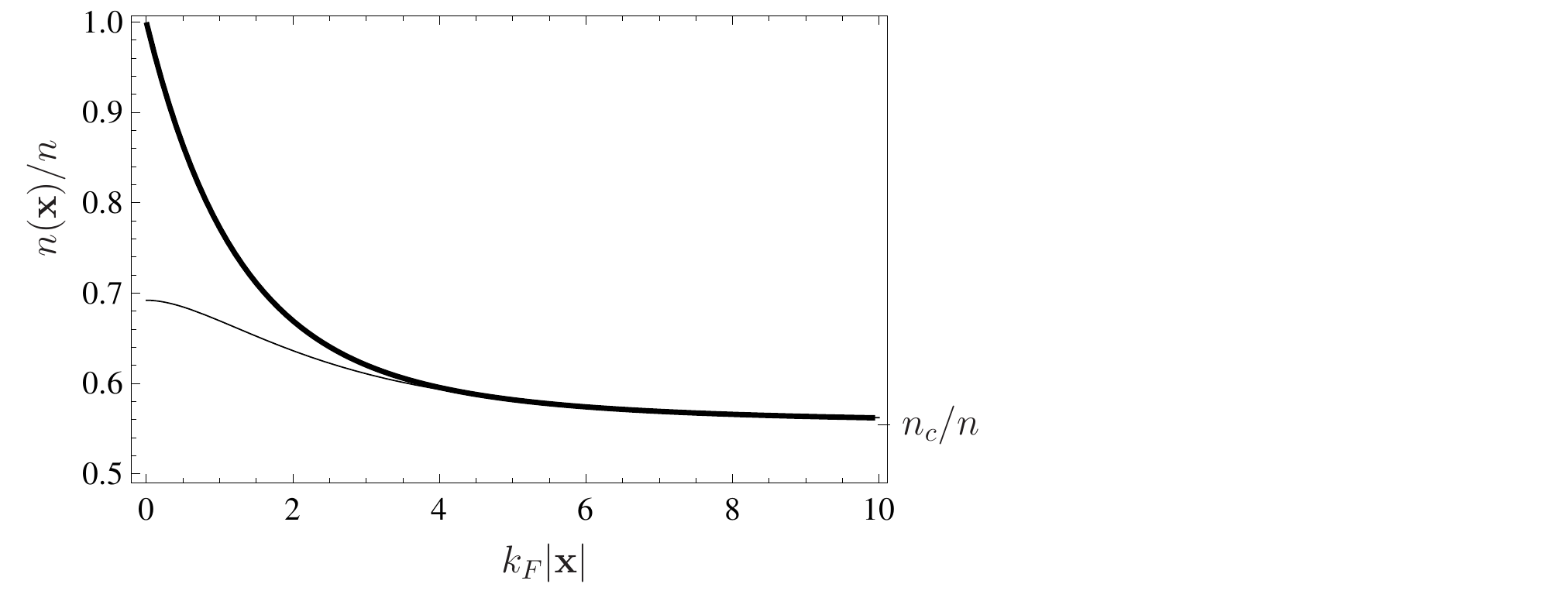}
\par\end{centering}

\caption{The universal one-particle density matrix $n(\mathbf{x})/n$ as a
function $k_{F}|\mathbf{x}|$, where the thin line is the contribution
of the condensate and its phase fluctuations as mentioned in the text.
From the difference of the graphs the contribution coming from all
other fluctuations can be inferred. The condensate density $n_{c}/n$
is indicated on the right.\label{fig:One_part_dens_matrix}}
\end{figure}
 In view of this possibility, there has been considerable theoretical
interest in the unitary Bose gas, but recent theoretical results strongly
vary from $\beta\simeq-0.34$ using renormalization-group techniques
\citep{Lee2010Universality}, $\beta\simeq1.93$ and $\beta\simeq-0.2$
from variational (Jastrow) analyses \citep{Cowell2002Cold,Song2009Ground},
to the complete instability of the Bose gas at unitarity using a Nozieres-Schmitt-Rink-like
approach \citep{Li2012Bose}. This clearly indicates that a systematic
approach for the meta-stable atomic Bose gas at unitarity is still
lacking.

In this Letter we present an analytical approach to the Bose gas at
unitarity that can be improved systematically. It is based on the
realization that fluctuations in the phase of the Bose-Einstein condensate
dominate the long-wavelength behavior of the system \citep{Popov1971Application,Popov1972Hydrodynamic,Gavoret1964Structure}.
Going beyond the Bogoliubov theory, which is required at unitarity,
these gapless phase fluctuations lead to infrared divergencies, which
make it increasingly difficult to apply resummation procedures to
find, for example, the effective interaction at unitarity. Our approach
circumvents these troublesome divergencies by exactly incorporating
the phase fluctuations, which will be confirmed by reproducing the
exact form of the single-particle propagator in the long-wavelength
limit as derived by Nepomnyashchii and Nepomnyashchii \citep{Nepomnyashchii1975Contribution,Nepomnyashchii1978Infrared}.
The approach thus consists of isolating the phase fluctuations of
the condensate, which is reminiscent of bosonization for fermions.
In addition, the theory is first renormalized due to all other fluctuations
using the renormalization group. One of the outcomes of our theory,
the universal one-particle density matrix, is shown in Fig.~\ref{fig:One_part_dens_matrix}.
The thin line indicates the contributions to the density matrix from
the condensate and its phase fluctuations, from which it is seen that
these fluctuations are vital to the correct description of the Bose
gas at unitarity.

\begin{figure}[t]
\begin{centering}
\includegraphics[scale=0.7]{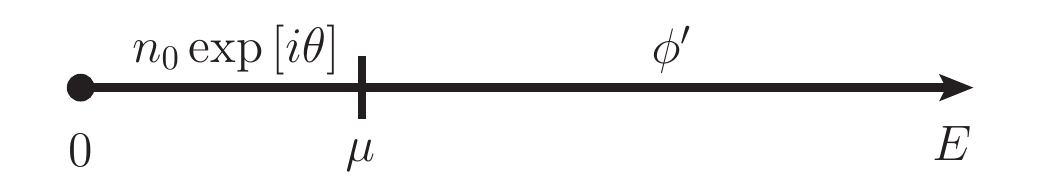}
\par\end{centering}

\caption{Schematic representation of the expansion of the field $\phi$ in
terms of the condensate and its phase fluctuations and the non-phase
fluctuations $\phi'$, c.f. Eq.~(\ref{eq:Expansion_of_field}).\label{fig:Expansion_wrt_scale}}
\end{figure}

\textit{Renormalized Bosonization.} --- The Euclidean action of a
Bose gas with a point interaction is given by $S\left[\phi^{*},\phi\right]=\int\mathrm{d}\tau\mathrm{d}\mathbf{x}\,\mathcal{L}$,
where the Lagrangian density is
\begin{equation}
\mathcal{L}=\phi^{*}\left[\hbar\partial_{\tau}-\frac{\hbar^{2}\nabla^{2}}{2m}-\mu\right]\phi+\frac{1}{2}T^{\mathrm{2B}}\left|\phi\right|^{4},\label{eq:Action_Bose_Gas}
\end{equation}
$\phi(\mathbf{x},\tau)$ is the bosonic field, and the strength of
the interaction is characterized by $T^{\mathrm{2B}}=4\pi a\hbar^{2}/m$
with $a$ the $s$-wave scattering length. To describe the Bose-Einstein
condensate at zero temperature, we expand the field as
\begin{equation}
\phi(\mathbf{x},\tau)=\sqrt{n_{0}}\exp\left[i\theta(\mathbf{x},\tau)\right]+\phi'(\mathbf{x},\tau),\label{eq:Expansion_of_field}
\end{equation}
where $n_{0}$ should be viewed as the quasicondensate density \citep{Stoof2009Ultracold}
and is not the density of atoms in the condensate $n_{c}$, as illustrated
by Fig.~\ref{fig:One_part_dens_matrix}. Roughly speaking, the first
term of the expansion describes the low-energy modes of the field,
as shown in Fig.~\ref{fig:Expansion_wrt_scale}, and includes the
phase fluctuations. The field $\phi'(\mathbf{x},\tau)$ describes
the high-energy modes and is defined such that it does not contain
phase fluctuations and is thus orthogonal to the first term in Eq.~(\ref{eq:Expansion_of_field}).
By inserting the expansion into Eq.~(\ref{eq:Action_Bose_Gas}),
the action $S\left[n_{0},\theta,\phi'^{*},\phi'\right]$ is obtained.
There are several properties of this action worth mentioning.

First of all, the dominant low-energy physics is known to be due to
the phase fluctuations. It can be shown that the exact form of the
phase-fluctuation action $S[\theta]$ in the long-wavelength limit
is
\begin{equation}
\frac{1}{2}\sum_{\mathbf{k},n}\theta^{*}(\mathbf{k},\omega_{n})\left[\frac{mc^{2}/n}{(\hbar\omega_{n})^{2}+2mc^{2}\epsilon_{\mathbf{k}}}\right]^{-1}\theta(\mathbf{k},\omega_{n}),\label{eq:Phase_fluc_prop}
\end{equation}
where $\omega_{n}=2n\pi/\hbar\beta$ is the Matsubara frequency,
$\beta^{-1}=k_{B}T$ is the temperature, $\mathbf{k}$ is the wave
vector, $\epsilon_{\mathbf{k}}=\hbar^{2}\mathbf{k}^{2}/2m$ is the
atomic dispersion, and $c$ is the speed of sound. In other words,
Eq.~(\ref{eq:Phase_fluc_prop}) can formally be obtained from the
full action $S\left[n_{0},\theta,\phi'^{*},\phi'\right]$ by integrating
out all non-phase fluctuations $\phi'$ and the fluctuations in $n_{0}$.

Second, the accuracy of the action $S\left[n_{0},\theta,\phi'^{*},\phi'\right]$
can be improved systematically by incorporating the $\phi'$ fluctuations
into a renormalization of the action. However, it turns out to be
more convenient to carry out this renormalization at the level of
$S\left[\phi^{*},\phi\right]$, Eq.~(\ref{eq:Action_Bose_Gas}),
and then apply the expansion of the field, as in Eq.~(\ref{eq:Expansion_of_field}).
The exact renormalization-group flow equation for the action $S\left[\phi^{*},\phi\right]$
is
\[
\frac{\mathrm{d}S}{\mathrm{d}\Lambda}=\frac{\hbar}{2}\mathrm{Tr}\delta_{\Lambda}\ln\left[-\mathbf{G}'^{-1}+\frac{1}{\hbar}\frac{\delta^{2}S_{\mathrm{int}}}{\delta\mathbf{\Phi}\delta\mathbf{\Phi}^{*}}\right].
\]
Here $S\left[\phi^{*},\phi;\Lambda\right]$ is the effective action
obtained by integrating out all non-phase fluctuations above the momentum
$\hbar\Lambda$, $\mathbf{G}'$ is the matrix propagator of the non-phase
fluctuations, $S{}_{\mathrm{int}}$ is the non-gaussian part of the
effective action, the trace is over space, imaginary time and Nambu
space $\mathbf{\Phi}(\mathbf{k},\omega_{n})=\left[\phi'(\mathbf{k},\omega_{n}),\phi'^{*}(-\mathbf{k},-\omega_{n})\right]^{T}$,
and $\delta_{\Lambda}=\delta(k-\Lambda)$. Although there are no small
parameters in the theory of unitary Bose gases, the renormalization
group can distinguish between the relevance of the various coupling
constants based on their scaling dimension under renormalization.
As the effective interaction evaluated at zero momentum and zero frequency
is expected to be a crucial variable, which however also induces a
flow of the chemical potential, let us here restrict our attention
to these parameters. The running of the chemical potential and effective
interaction are in general found to be given by $\Lambda\mathrm{d}\mu/\mathrm{d}\Lambda=\beta_{\mu}(\mu,g)$
and $\Lambda\mathrm{d}g/\mathrm{d}\Lambda=\beta_{g}(\mu,g)$. By solving
these equations the renormalized action $S\left[\phi^{*},\phi;\Lambda\right]$
is found. Then, after inserting the expansion of the field, the renormalized
action $S\left[n_{0},\theta,\phi'^{*},\phi';\Lambda\right]$ is obtained.
This action defines the propagator of the non-phase fluctuations in
terms of the effective interaction.

Lastly, to actually perform the above-mentioned renormalization, the
propagator of the non-phase fluctuations $\phi'$ is needed. It is
obtained by realizing that these fluctuations are determined from
the action $S\left[n_{0},\theta,\phi'^{*},\phi';\Lambda\right]$ with
a non-fluctuating phase. The quadratic part of the action is then
written as $S\left[\phi'^{*},\phi'\right]=\frac{1}{2}\sum_{\mathbf{k},n}\mathbf{\Phi}^{\dagger}(\mathbf{k},\omega_{n})\left[-\hbar\mathbf{G}'^{-1}(\mathbf{k},\omega_{n})\right]\mathbf{\Phi}(\mathbf{k},\omega_{n})$,
where the components of the symmetric $2\times2$ Green's function
of the non-phase fluctuations are given by
\begin{align*}
-G'_{11}(\mathbf{k},\omega_{n}) & =\hbar\frac{i\hbar\omega_{n}+\epsilon_{\mathbf{k}}+n_{0}g}{(\hbar\omega_{n})^{2}+(\hbar\omega_{\mathbf{k}})^{2}}-n_{0}\left\langle \theta\theta^{*}\right\rangle (\mathbf{k},\omega_{n}),\\
-G'_{12}(\mathbf{k},\omega_{n}) & =\hbar\frac{-n_{0}g}{(\hbar\omega_{n})^{2}+(\hbar\omega_{\mathbf{k}})^{2}}+n_{0}\left\langle \theta\theta^{*}\right\rangle (\mathbf{k},\omega_{n}),
\end{align*}
with the Bogoliubov dispersion $(\hbar\omega_{\mathbf{k}})^{2}=\epsilon_{\mathbf{k}}(\epsilon_{\mathbf{k}}+2n_{0}g)$.

The first term in the propagator is the standard Bogoliubov propagator,
which also includes the effect of phase fluctuations. The second term
precisely removes these effects, which can for instance be identified
by their proportionality to $n_{0}$ in the numerator. For consistency,
we define $mc^{2}\equiv n_{0}g$ in accordance with the Bogoliubov
dispersion. Notice, however, that the residues of the exact propagator
of the phase fluctuations incorporate the renormalization of the superfluid
density from $n_{0}$ to $n$, which is not present in the Bogoliubov
propagator.

After the subtraction of the phase fluctuations, the propagator of
the non-phase fluctuations is given by
\begin{equation}
\hbar^{-1}\left\langle \phi'(\mathbf{k},\omega_{n})\phi'^{*}(\mathbf{k},\omega_{n})\right\rangle =\frac{i\hbar\omega_{n}+\epsilon_{\mathbf{k}}}{(\hbar\omega_{n})^{2}+(\hbar\omega_{\mathbf{k}})^{2}},\label{eq:High_En_fluc}
\end{equation}
and the anomalous averages vanish, i.e., $\left\langle \phi'\phi'\right\rangle =\left\langle \phi'^{*}\phi'^{*}\right\rangle =0$.
The vanishing of the anomalous averages means that the Green's function
$\mathbf{G}'$ is diagonal and this greatly simplifies the renormalization
procedure of the interaction, which justifies our preference for solving
the renormalization-group equation for $S\left[\phi^{*},\phi\right]$.

Before we turn to the determination of the effective interaction,
we first show that our approach reproduces the exact propagator in
the static long-wavelength limit derived by Nepomnyashchii and Nepomnyashchii,
as mentioned in the introduction. This is achieved using the one-particle
correlation function $\left\langle \phi(\mathbf{x},\tau)\phi^{*}(\mathbf{0},0)\right\rangle $,
which equals $n_{0}\left\langle \exp\left[i\left(\theta(\mathbf{x},\tau)-\theta(\mathbf{0},0)\right)\right]\right\rangle +\left\langle \phi'(\mathbf{x},\tau)\phi'^{*}(\mathbf{0},0)\right\rangle $,
and taking the Fourier transform. By expanding the exponential, we
find that the dominant long-wavelength behavior is due only to the
first two terms in the expansion, where the first term is simply the
exact phase-fluctuation propagator in Eq.~(\ref{eq:Phase_fluc_prop})
while the second term comes from a convolution of two propagators,
namely
\begin{align*}
 & \hbar^{-1}\left\langle \phi(\mathbf{k},\omega_{n})\phi{}^{*}(\mathbf{k},\omega_{n})\right\rangle \simeq\frac{\frac{n_{c}}{n}mc^{2}}{(\hbar\omega_{n})^{2}+2mc^{2}\epsilon_{\mathbf{k}}}\\
 & -\frac{3\sqrt{mc^{2}}}{32\sqrt{2}\epsilon_{F}^{3/2}}\frac{n_{c}}{n}\log\left[\frac{(\hbar\omega_{n})^{2}+2mc^{2}\epsilon_{\mathbf{k}}}{(8mc^{2})^{2}}\right]+n_{c}\beta V\delta_{\mathbf{k},\mathbf{0}}\delta_{n,0},
\end{align*}
which is the form of the exact propagator \citep{Nepomnyashchii1975Contribution,Nepomnyashchii1978Infrared}
with $n_{c}$ the condensate density defined later in Eq.~(\ref{eq:cond_density})
and $V$ is the volume. The exact anomalous propagator is also reproduced,
and is given by $n_{0}\left\langle \exp\left[i\left(\theta(\mathbf{x},\tau)+\theta(\mathbf{0},0)\right)\right]\right\rangle $.
In particular, this leads to the counter-intuitive conclusion that
the anomalous self-energy vanishes at zero momentum and frequency
\citep{Nepomnyashchii1978Infrared}.

\begin{figure}[t]
\begin{centering}
\includegraphics[scale=0.5]{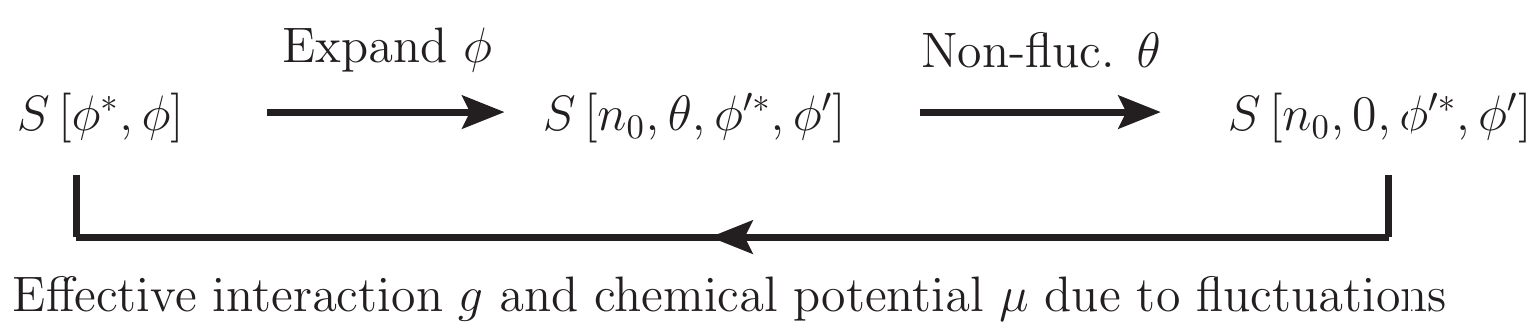}
\par\end{centering}

\caption{Schematic representation of the renormalization procedure that shows
the self-consistent nature of our theory, as described in the text.\label{fig:theory-diagram}}
\end{figure}

To summarize and as illustrated in Fig.~\ref{fig:theory-diagram},
the action $S\left[\phi{}^{*},\phi\right]$ of the Bose gas can be
systematically renormalized by the non-phase fluctuations $\phi'$
using the renormalization-group flow equation, giving in particular
rise to an effective coupling $g$ and a renormalized chemical potential
$\mu$. The propagators of the non-phase fluctuations are determined
self-consistently after expansion of the field. The theory includes
the exact propagator of phase fluctuations and can reproduce the exact
propagator $\left\langle \phi\phi^{*}\right\rangle $ in the long-wavelength
limit.

\textit{Universal Results.} --- Now that our theoretical framework
has been established it is possible to determine, for example, the
condensate density, the effective interaction and the chemical potential
at unitarity. The condensate density can be calculated from the off-diagonal
long-range order, namely $n_{c}\equiv\lim_{|\mathbf{x}|\rightarrow\infty}\left\langle \phi(\mathbf{x},0)\phi^{*}(\mathbf{0},0)\right\rangle $,
\begin{equation}
n_{c}=n_{0}\exp\left[\frac{3}{4}\left(2\sqrt{2}-\pi\right)\left(\frac{n_{0}g}{\frac{\hbar^{2}}{2m}(6\pi^{2}n_{0})^{2/3}}\right)^{3/2}\right],\label{eq:cond_density}
\end{equation}
where it must be noted that an ultra-violet subtraction is needed
in order to calculate the expectation value of the phase fluctuations
using Eq.~(\ref{eq:Phase_fluc_prop}) with $mc^{2}=n_{0}g$ \citep{Stoof2009Ultracold}.
This ultra-violet subtraction removes the ultra-violet divergences
associated with a point interaction, and is a result of the renormalization
of the bare coupling to $T^{\mathrm{2B}}$. In order to determine
the condensate density at unitarity, the quasicondensate density $n_{0}$
needs to be eliminated in favor of the total density $n=\left\langle \phi(\mathbf{x},\tau)\phi^{*}(\mathbf{x},\tau)\right\rangle $
using
\begin{equation}
n=n_{0}+\frac{\left(8\sqrt{2}-3\pi\right)}{24\pi^{2}}\left(\frac{2m}{\hbar^{2}}\left[n_{0}g\right]\right)^{3/2},\label{eq:Total_density}
\end{equation}
where the second term is the contribution from the high-energy fluctuations
$n'=\left\langle \phi'(\mathbf{x},\tau)\phi'^{*}(\mathbf{x},\tau)\right\rangle $,
see Eq.~(\ref{eq:High_En_fluc}). As required, exactly the same ultra-violet
subtraction was used for the high-energy fluctuations as in Eq.~(\ref{eq:cond_density}).

To proceed, we must now determine the effective interaction $g$ in
some approximation. Taking only the renormalization of the coupling
constant and the chemical potential into account, which turns out
to be very accurate for the unitary Fermi gas \citep{Gubbels2008Renormalization},
the beta functions are given by
\begin{figure}[b]
\centering{}\includegraphics[scale=0.35]{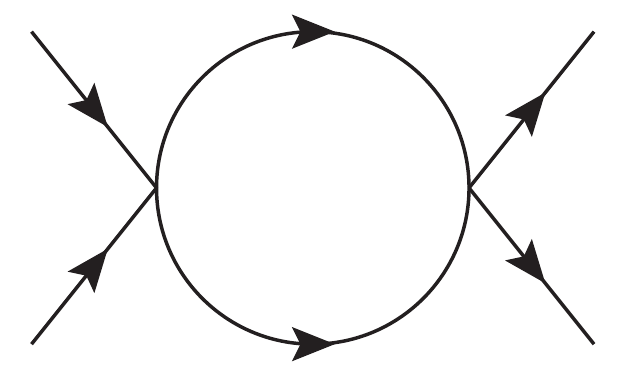}\qquad{}\qquad{}\includegraphics[scale=0.35]{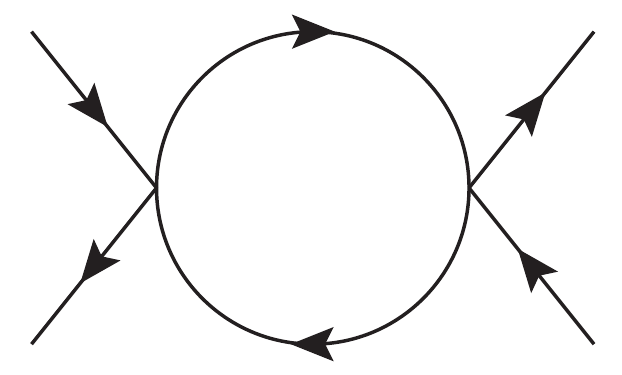}\caption{The Feynman diagrams for the fluctuations $\phi'$ that contribute
to the beta function $\beta_{g}$ and can be viewed as the building
blocks for the ladder and bubble sums included in the Bethe-Salpeter
equation for the effective interaction $g$.\label{fig:loop_bubble_diagram}}
\end{figure}
\begin{align}
\beta_{\mu} & =-2g\frac{4\pi\Lambda^{3}}{(2\pi)^{3}}\left[|v_{\Lambda}|^{2}+\frac{n_{0}g}{2\epsilon_{\Lambda}+2n_{0}g}\right],\label{eq:Beta_func}\\
\beta_{g} & =g^{2}\frac{4\pi\Lambda^{3}}{(2\pi)^{3}}\left[\frac{|u_{\Lambda}|^{4}+|v_{\Lambda}|^{4}-8|u_{\Lambda}|^{2}|v_{\Lambda}|^{2}}{2\hbar\omega_{\Lambda}}-\frac{1}{2\epsilon_{\Lambda}}\right],\nonumber
\end{align}
where the Bogoliubov dispersion $\hbar\omega_{\mathbf{k}}$ and the
coherence factors $|u_{\mathbf{k}}|^{2}=|v_{\mathbf{k}}|^{2}+1=\left(\hbar\omega_{\mathbf{k}}+\epsilon_{\mathbf{k}}\right)/2\hbar\omega_{\mathbf{k}}$
are evaluated at $\Lambda$. The effective interaction is obtained
by integrating its differential equation using the boundary condition
$g(\Lambda=\infty)=T^{\mathrm{2B}}$, where it must be noted that
the effective interaction inside the Bogoliubov dispersion is the
fully renormalized value $g(\Lambda=0)$ which, as previously explained,
is determined self-consistently. Ultimately, we obtain
\begin{figure}[t]
\begin{centering}
\includegraphics[bb=15bp 10bp 345bp 235bp,clip,scale=0.7]{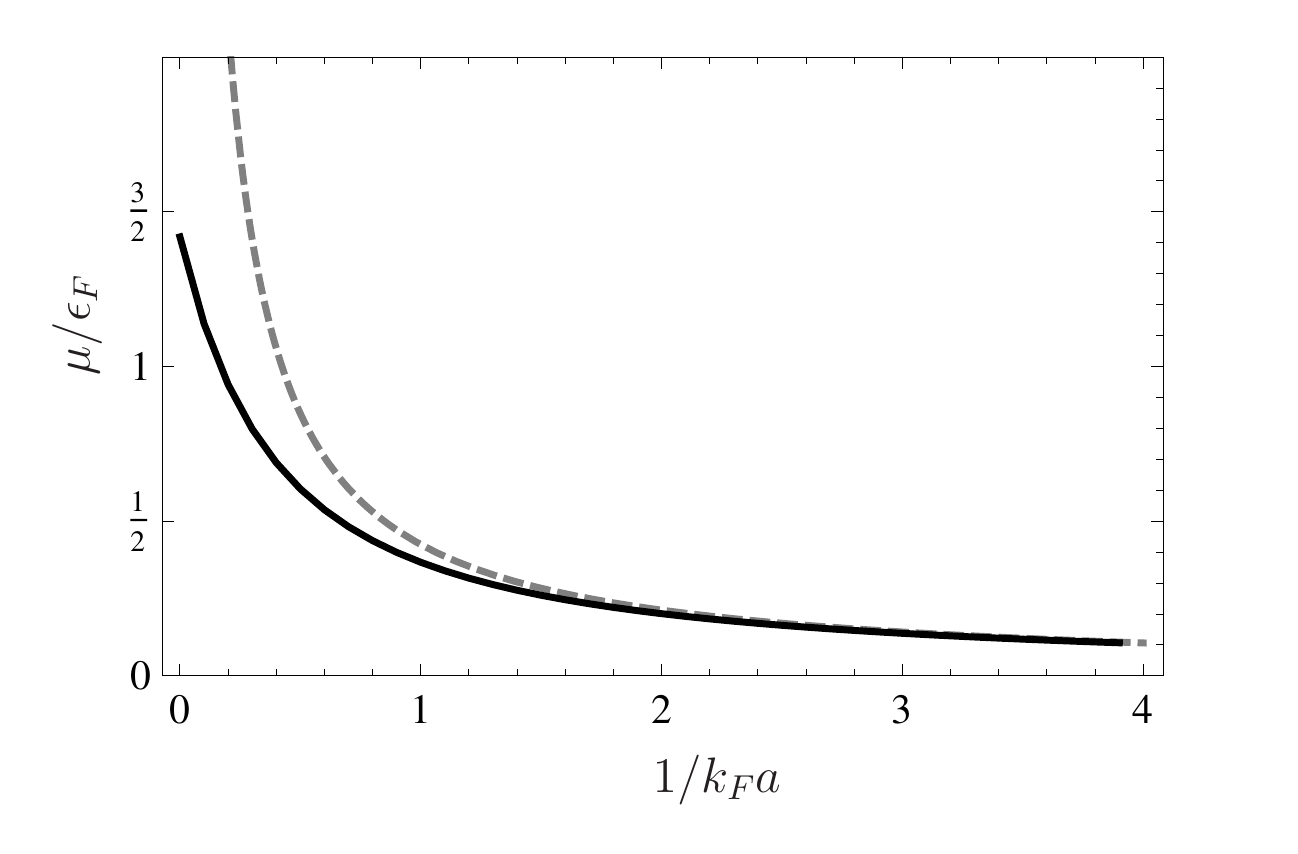}
\par\end{centering}

\caption{The chemical potential as a function of scattering length. The gray
dashed line is the Bogoliubov chemical potential \citep{Stoof2009Ultracold}.\label{fig:Chem_pot_vs_scat_length}}
\end{figure}
\begin{equation}
\frac{1}{g}=\frac{1}{T^{\mathrm{2B}}}+\frac{1}{4\sqrt{2}\pi^{2}}\left(\frac{2m}{\hbar^{2}}\right)^{3/2}\sqrt{n_{0}g}.\label{eq:Eff_int2}
\end{equation}
Note that this equation can also be obtained directly as the result
of a resummation of an infinite number of the diagrams shown in Fig.~\ref{fig:loop_bubble_diagram}.
The equation also shows the highly non-perturbative nature of the
renormalization group.

In the unitarity limit, $T^{2B}\rightarrow\infty$, the effective
interaction and the condensate density in terms of the total density
are found by solving Eqs.~(\ref{eq:cond_density}-\ref{eq:Eff_int2})
to be
\begin{align*}
\frac{ng}{\epsilon_{F}} & =\frac{2}{3^{2/3}}\left(1+\lambda\right)^{1/3}\simeq1.09,\quad\frac{n'}{n}=\frac{\lambda}{1+\lambda}\simeq0.31,\\
\frac{n_{c}}{n} & =\frac{1}{1+\lambda}\exp\left(\frac{2\sqrt{2}-\pi}{\sqrt{2}}\right)\simeq0.55,
\end{align*}
where $\lambda=n'/n_{0}=\left(8\sqrt{2}-3\pi\right)/3\sqrt{2}\simeq0.45$.
The depletion from the condensate is given by $1-n_{c}/n\simeq0.45$,
which clearly differs from the density of particles contributing to
the non-phase fluctuating modes $n'$ by phase-fluctuation contributions.

The universal one-particle density matrix $n(\mathbf{x})=\left\langle \phi(\mathbf{x},0)\phi^{*}(\mathbf{0},0)\right\rangle $
is shown in Fig.~\ref{fig:One_part_dens_matrix} and can be seen
to reduce in the large-distance limit to the condensate density. Also
the contribution due to the phase fluctuations is indicated, which
terminates at $n_{0}/n$ at equal position. Another interesting property
at unitarity is called the contact $C$ and is related to the short-wavelength
behavior of the one-particle density matrix, namely $n(\mathbf{k})\simeq C/\mathbf{k}^{4}$
\citep{Tan2008Large,Braaten2011Universal}. The contact is
\[
\frac{C}{k_{F}^{4}}=\left(\frac{n_{0}g}{2\epsilon_{F}}\right)^{2}=\frac{1}{3^{4/3}}\frac{1}{\left(1+\lambda\right)^{4/3}}\simeq0.14.
\]
The change in the chemical potential follows from integrating Eq.
(\ref{eq:Beta_func}) and is given by $\Delta\mu=2n'g$. According
to the exact Hugenholtz-Pines theorem the chemical potential in our
theory is then given by $\mu=n_{0}g+\Delta\mu$, such that the universal
chemical potential is
\[
\frac{\mu}{\epsilon_{F}}=\frac{n_{0}g+2n'g}{\epsilon_{F}}=\frac{2}{3^{2/3}}\frac{1+2\lambda}{\left(1+\lambda\right)^{2/3}}\simeq1.42,
\]
which results in the universal constant $\beta\simeq0.42$. Furthermore,
the speed of sound at unitarity is given by
\[
\frac{mc^{2}}{\epsilon_{F}}=\frac{n_{0}g}{\epsilon_{F}}=\frac{1}{1+2\lambda}\frac{\mu}{\epsilon_{F}}\simeq0.53\frac{\mu}{\epsilon_{F}}\simeq0.75.
\]
The expected value for the speed of sound at unitarity in terms of
the chemical potential is $mc^{2}=n(\mathrm{d}\mu/\mathrm{d}n)=2\mu/3\simeq0.66\mu$,
which is remarkably close to our result and shows the accuracy of
the simplest first approximation that we have presented here.

\textit{Conclusions.} --- We have constructed an approach to describe
Bose gases at unitarity which can be improved systematically by renormalization-group
methods. As an outlook, all quantities can be found not only at unitarity,
but it is possible using Eq.~(\ref{eq:Eff_int2}) to find all these
quantities from weak to infinitely strong coupling, as is shown for
the chemical potential in Fig.~\ref{fig:Chem_pot_vs_scat_length}.
The generalization of the theory to frequency-dependent interactions
and also non-zero temperature is straightforward and will be elaborated
on in a future publication. Furthermore, we expect that the approach
can be applied to other systems with a broken continuous symmetry,
where similar infrared divergencies occur as a consequence of the
Goldstone modes. We hope that our results stimulate further experimental
developments toward unitarity-limited Bose gases in the near future.
\begin{acknowledgments}
This work is supported by the Stichting voor Fundamenteel Onderzoek
der Materie (FOM) and the Nederlandse Organisatie voor Wetenschaplijk
Onderzoek (NWO).
\end{acknowledgments}
\bibliographystyle{apsrev4-1}
\bibliography{article}

\begin{thebibliography}{18}%
\makeatletter
\providecommand \@ifxundefined [1]{%
 \@ifx{#1\undefined}
}%
\providecommand \@ifnum [1]{%
 \ifnum #1\expandafter \@firstoftwo
 \else \expandafter \@secondoftwo
 \fi
}%
\providecommand \@ifx [1]{%
 \ifx #1\expandafter \@firstoftwo
 \else \expandafter \@secondoftwo
 \fi
}%
\providecommand \natexlab [1]{#1}%
\providecommand \enquote  [1]{``#1''}%
\providecommand \bibnamefont  [1]{#1}%
\providecommand \bibfnamefont [1]{#1}%
\providecommand \citenamefont [1]{#1}%
\providecommand \href@noop [0]{\@secondoftwo}%
\providecommand \href [0]{\begingroup \@sanitize@url \@href}%
\providecommand \@href[1]{\@@startlink{#1}\@@href}%
\providecommand \@@href[1]{\endgroup#1\@@endlink}%
\providecommand \@sanitize@url [0]{\catcode `\\12\catcode `\$12\catcode
  `\&12\catcode `\#12\catcode `\^12\catcode `\_12\catcode `\%12\relax}%
\providecommand \@@startlink[1]{}%
\providecommand \@@endlink[0]{}%
\providecommand \url  [0]{\begingroup\@sanitize@url \@url }%
\providecommand \@url [1]{\endgroup\@href {#1}{\urlprefix }}%
\providecommand \urlprefix  [0]{URL }%
\providecommand \Eprint [0]{\href }%
\providecommand \doibase [0]{http://dx.doi.org/}%
\providecommand \selectlanguage [0]{\@gobble}%
\providecommand \bibinfo  [0]{\@secondoftwo}%
\providecommand \bibfield  [0]{\@secondoftwo}%
\providecommand \translation [1]{[#1]}%
\providecommand \BibitemOpen [0]{}%
\providecommand \bibitemStop [0]{}%
\providecommand \bibitemNoStop [0]{.\EOS\space}%
\providecommand \EOS [0]{\spacefactor3000\relax}%
\providecommand \BibitemShut  [1]{\csname bibitem#1\endcsname}%
\let\auto@bib@innerbib\@empty
\bibitem [{\citenamefont {Bloch}\ \emph {et~al.}(2008)\citenamefont {Bloch},
  \citenamefont {Dalibard},\ and\ \citenamefont {Zwerger}}]{Bloch2008Manybody}%
  \BibitemOpen
  \bibfield  {author} {\bibinfo {author} {\bibfnamefont {I.}~\bibnamefont
  {Bloch}}, \bibinfo {author} {\bibfnamefont {J.}~\bibnamefont {Dalibard}}, \
  and\ \bibinfo {author} {\bibfnamefont {W.}~\bibnamefont {Zwerger}},\ }\href
  {\doibase 10.1103/RevModPhys.80.885} {\bibfield  {journal} {\bibinfo
  {journal} {Rev. Mod. Phys.}\ }\textbf {\bibinfo {volume} {80}},\ \bibinfo
  {pages} {885} (\bibinfo {year} {2008})}\BibitemShut {NoStop}%
\bibitem [{\citenamefont {Chin}\ \emph {et~al.}(2010)\citenamefont {Chin},
  \citenamefont {Grimm}, \citenamefont {Julienne},\ and\ \citenamefont
  {Tiesinga}}]{Chin2010Feshbach}%
  \BibitemOpen
  \bibfield  {author} {\bibinfo {author} {\bibfnamefont {C.}~\bibnamefont
  {Chin}}, \bibinfo {author} {\bibfnamefont {R.}~\bibnamefont {Grimm}},
  \bibinfo {author} {\bibfnamefont {P.}~\bibnamefont {Julienne}}, \ and\
  \bibinfo {author} {\bibfnamefont {E.}~\bibnamefont {Tiesinga}},\ }\href
  {http://dx.doi.org/10.1103/RevModPhys.82.1225} {\bibfield  {journal}
  {\bibinfo  {journal} {Rev. Mod. Phys.}\ }\textbf {\bibinfo {volume} {82}},\
  \bibinfo {pages} {1225} (\bibinfo {year} {2010})}\BibitemShut {NoStop}%
\bibitem [{\citenamefont {Stoof}\ \emph {et~al.}(2009)\citenamefont {Stoof},
  \citenamefont {Dickerscheid},\ and\ \citenamefont
  {Gubbels}}]{Stoof2009Ultracold}%
  \BibitemOpen
  \bibfield  {author} {\bibinfo {author} {\bibfnamefont {H.~T.~C.}\
  \bibnamefont {Stoof}}, \bibinfo {author} {\bibfnamefont {D.~B.~M.}\
  \bibnamefont {Dickerscheid}}, \ and\ \bibinfo {author} {\bibfnamefont
  {K.}~\bibnamefont {Gubbels}},\ }\href
  {http://www.worldcat.org/isbn/1402087624} {\emph {\bibinfo {title} {Ultracold
  Quantum Fields (Theoretical and Mathematical Physics)}}},\ \bibinfo {edition}
  {1st}\ ed.\ (\bibinfo  {publisher} {Springer},\ \bibinfo {year}
  {2009})\BibitemShut {NoStop}%
\bibitem [{\citenamefont {Ku}\ \emph {et~al.}(2012)\citenamefont {Ku},
  \citenamefont {Sommer}, \citenamefont {Cheuk},\ and\ \citenamefont
  {Zwierlein}}]{Ku2012Revealing}%
  \BibitemOpen
  \bibfield  {author} {\bibinfo {author} {\bibfnamefont {M.~J.~H.}\
  \bibnamefont {Ku}}, \bibinfo {author} {\bibfnamefont {A.~T.}\ \bibnamefont
  {Sommer}}, \bibinfo {author} {\bibfnamefont {L.~W.}\ \bibnamefont {Cheuk}}, \
  and\ \bibinfo {author} {\bibfnamefont {M.~W.}\ \bibnamefont {Zwierlein}},\
  }\href {\doibase 10.1126/science.1214987} {\bibfield  {journal} {\bibinfo
  {journal} {Science}\ }\textbf {\bibinfo {volume} {335}},\ \bibinfo {pages}
  {563} (\bibinfo {year} {2012})}\BibitemShut {NoStop}%
\bibitem [{\citenamefont {Z\"{u}rn}\ \emph {et~al.}(2012)\citenamefont
  {Z\"{u}rn}, \citenamefont {Lompe}, \citenamefont {Wenz}, \citenamefont
  {Jochim}, \citenamefont {Julienne},\ and\ \citenamefont
  {Hutson}}]{Zurn2012Precise}%
  \BibitemOpen
  \bibfield  {author} {\bibinfo {author} {\bibfnamefont {G.}~\bibnamefont
  {Z\"{u}rn}}, \bibinfo {author} {\bibfnamefont {T.}~\bibnamefont {Lompe}},
  \bibinfo {author} {\bibfnamefont {A.~N.}\ \bibnamefont {Wenz}}, \bibinfo
  {author} {\bibfnamefont {S.}~\bibnamefont {Jochim}}, \bibinfo {author}
  {\bibfnamefont {P.~S.}\ \bibnamefont {Julienne}}, \ and\ \bibinfo {author}
  {\bibfnamefont {J.~M.}\ \bibnamefont {Hutson}},\ }\href
  {http://arxiv.org/abs/1211.1512} {\  (\bibinfo {year} {2012})},\ \Eprint
  {http://arxiv.org/abs/1211.1512} {arXiv:1211.1512} \BibitemShut {NoStop}%
\bibitem [{\citenamefont {Navon}\ \emph {et~al.}(2011)\citenamefont {Navon},
  \citenamefont {Piatecki}, \citenamefont {G\"{u}nter}, \citenamefont {Rem},
  \citenamefont {Nguyen}, \citenamefont {Chevy}, \citenamefont {Krauth},\ and\
  \citenamefont {Salomon}}]{Navon2011Dynamics}%
  \BibitemOpen
  \bibfield  {author} {\bibinfo {author} {\bibfnamefont {N.}~\bibnamefont
  {Navon}}, \bibinfo {author} {\bibfnamefont {S.}~\bibnamefont {Piatecki}},
  \bibinfo {author} {\bibfnamefont {K.}~\bibnamefont {G\"{u}nter}}, \bibinfo
  {author} {\bibfnamefont {B.}~\bibnamefont {Rem}}, \bibinfo {author}
  {\bibfnamefont {T.~C.}\ \bibnamefont {Nguyen}}, \bibinfo {author}
  {\bibfnamefont {F.}~\bibnamefont {Chevy}}, \bibinfo {author} {\bibfnamefont
  {W.}~\bibnamefont {Krauth}}, \ and\ \bibinfo {author} {\bibfnamefont
  {C.}~\bibnamefont {Salomon}},\ }\href {\doibase
  10.1103/PhysRevLett.107.135301} {\bibfield  {journal} {\bibinfo  {journal}
  {Phys. Rev. Lett.}\ }\textbf {\bibinfo {volume} {107}},\ \bibinfo {pages}
  {135301} (\bibinfo {year} {2011})}\BibitemShut {NoStop}%
\bibitem [{\citenamefont {Lee}\ and\ \citenamefont
  {Lee}(2010)}]{Lee2010Universality}%
  \BibitemOpen
  \bibfield  {author} {\bibinfo {author} {\bibfnamefont {Y.~L.}\ \bibnamefont
  {Lee}}\ and\ \bibinfo {author} {\bibfnamefont {Y.~W.}\ \bibnamefont {Lee}},\
  }\href {\doibase 10.1103/PhysRevA.81.063613} {\bibfield  {journal} {\bibinfo
  {journal} {Phys. Rev. A}\ }\textbf {\bibinfo {volume} {81}},\ \bibinfo
  {pages} {063613} (\bibinfo {year} {2010})}\BibitemShut {NoStop}%
\bibitem [{\citenamefont {Cowell}\ \emph {et~al.}(2002)\citenamefont {Cowell},
  \citenamefont {Heiselberg}, \citenamefont {Mazets}, \citenamefont {Morales},
  \citenamefont {Pandharipande},\ and\ \citenamefont
  {Pethick}}]{Cowell2002Cold}%
  \BibitemOpen
  \bibfield  {author} {\bibinfo {author} {\bibfnamefont {S.}~\bibnamefont
  {Cowell}}, \bibinfo {author} {\bibfnamefont {H.}~\bibnamefont {Heiselberg}},
  \bibinfo {author} {\bibfnamefont {I.~E.}\ \bibnamefont {Mazets}}, \bibinfo
  {author} {\bibfnamefont {J.}~\bibnamefont {Morales}}, \bibinfo {author}
  {\bibfnamefont {V.~R.}\ \bibnamefont {Pandharipande}}, \ and\ \bibinfo
  {author} {\bibfnamefont {C.~J.}\ \bibnamefont {Pethick}},\ }\href {\doibase
  10.1103/PhysRevLett.88.210403} {\bibfield  {journal} {\bibinfo  {journal}
  {Phys. Rev. Lett.}\ }\textbf {\bibinfo {volume} {88}},\ \bibinfo {pages}
  {210403} (\bibinfo {year} {2002})}\BibitemShut {NoStop}%
\bibitem [{\citenamefont {Song}\ and\ \citenamefont
  {Zhou}(2009)}]{Song2009Ground}%
  \BibitemOpen
  \bibfield  {author} {\bibinfo {author} {\bibfnamefont {J.~L.}\ \bibnamefont
  {Song}}\ and\ \bibinfo {author} {\bibfnamefont {F.}~\bibnamefont {Zhou}},\
  }\href {\doibase 10.1103/PhysRevLett.103.025302} {\bibfield  {journal}
  {\bibinfo  {journal} {Phys. Rev. Lett.}\ }\textbf {\bibinfo {volume} {103}},\
  \bibinfo {pages} {025302} (\bibinfo {year} {2009})}\BibitemShut {NoStop}%
\bibitem [{\citenamefont {Li}\ and\ \citenamefont {Ho}(2012)}]{Li2012Bose}%
  \BibitemOpen
  \bibfield  {author} {\bibinfo {author} {\bibfnamefont {W.}~\bibnamefont
  {Li}}\ and\ \bibinfo {author} {\bibfnamefont {T.~L.}\ \bibnamefont {Ho}},\
  }\href {\doibase 10.1103/PhysRevLett.108.195301} {\bibfield  {journal}
  {\bibinfo  {journal} {Phys. Rev. Lett.}\ }\textbf {\bibinfo {volume} {108}},\
  \bibinfo {pages} {195301} (\bibinfo {year} {2012})}\BibitemShut {NoStop}%
\bibitem [{\citenamefont {Popov}(1971)}]{Popov1971Application}%
  \BibitemOpen
  \bibfield  {author} {\bibinfo {author} {\bibfnamefont {V.~N.}\ \bibnamefont
  {Popov}},\ }\href {\doibase 10.1007/BF01037581} {\bibfield  {journal}
  {\bibinfo  {journal} {Theor. Math. Phys.}\ }\textbf {\bibinfo {volume} {6}},\
  \bibinfo {pages} {65} (\bibinfo {year} {1971})}\BibitemShut {NoStop}%
\bibitem [{\citenamefont {Popov}(1972)}]{Popov1972Hydrodynamic}%
  \BibitemOpen
  \bibfield  {author} {\bibinfo {author} {\bibfnamefont {V.~N.}\ \bibnamefont
  {Popov}},\ }\href {\doibase 10.1007/BF01028563} {\bibfield  {journal}
  {\bibinfo  {journal} {Theor. Math. Phys.}\ }\textbf {\bibinfo {volume}
  {11}},\ \bibinfo {pages} {478} (\bibinfo {year} {1972})}\BibitemShut
  {NoStop}%
\bibitem [{\citenamefont {Gavoret}\ and\ \citenamefont
  {Nozi\`{e}res}(1964)}]{Gavoret1964Structure}%
  \BibitemOpen
  \bibfield  {author} {\bibinfo {author} {\bibfnamefont {J.}~\bibnamefont
  {Gavoret}}\ and\ \bibinfo {author} {\bibfnamefont {P.}~\bibnamefont
  {Nozi\`{e}res}},\ }\href {\doibase 10.1016/0003-4916(64)90200-3} {\bibfield
  {journal} {\bibinfo  {journal} {Ann. Phys.}\ }\textbf {\bibinfo {volume}
  {28}},\ \bibinfo {pages} {349} (\bibinfo {year} {1964})}\BibitemShut
  {NoStop}%
\bibitem [{\citenamefont {Nepomnyashchii}\ and\ \citenamefont
  {Nepomnyashchii}(1975)}]{Nepomnyashchii1975Contribution}%
  \BibitemOpen
  \bibfield  {author} {\bibinfo {author} {\bibfnamefont {A.~A.}\ \bibnamefont
  {Nepomnyashchii}}\ and\ \bibinfo {author} {\bibfnamefont {Y.}~\bibnamefont
  {Nepomnyashchii}},\ }\href
  {http://www.jetpletters.ac.ru/ps/1460/article_22243.shtml} {\bibfield
  {journal} {\bibinfo  {journal} {JETP Lett.}\ }\textbf {\bibinfo {volume}
  {21}},\ \bibinfo {pages} {3} (\bibinfo {year} {1975})}\BibitemShut {NoStop}%
\bibitem [{\citenamefont {Nepomnyashchii}\ and\ \citenamefont
  {Nepomnyashchii}(1978)}]{Nepomnyashchii1978Infrared}%
  \BibitemOpen
  \bibfield  {author} {\bibinfo {author} {\bibfnamefont {Y.}~\bibnamefont
  {Nepomnyashchii}}\ and\ \bibinfo {author} {\bibfnamefont {A.~A.}\
  \bibnamefont {Nepomnyashchii}},\ }\href
  {http://jetp.ac.ru/cgi-bin/e/index/e/48/3/p493?a=list} {\bibfield  {journal}
  {\bibinfo  {journal} {JETP}\ }\textbf {\bibinfo {volume} {48}},\ \bibinfo
  {pages} {493} (\bibinfo {year} {1978})}\BibitemShut {NoStop}%
\bibitem [{\citenamefont {Gubbels}\ and\ \citenamefont
  {Stoof}(2008)}]{Gubbels2008Renormalization}%
  \BibitemOpen
  \bibfield  {author} {\bibinfo {author} {\bibfnamefont {K.~B.}\ \bibnamefont
  {Gubbels}}\ and\ \bibinfo {author} {\bibfnamefont {H.~T.~C.}\ \bibnamefont
  {Stoof}},\ }\href {\doibase 10.1103/physrevlett.100.140407} {\bibfield
  {journal} {\bibinfo  {journal} {Physical Review Letters}\ }\textbf {\bibinfo
  {volume} {100}},\ \bibinfo {pages} {140407} (\bibinfo {year}
  {2008})}\BibitemShut {NoStop}%
\bibitem [{\citenamefont {Tan}(2008)}]{Tan2008Large}%
  \BibitemOpen
  \bibfield  {author} {\bibinfo {author} {\bibfnamefont {S.}~\bibnamefont
  {Tan}},\ }\href {http://dx.doi.org/10.1016/j.aop.2008.03.005} {\bibfield
  {journal} {\bibinfo  {journal} {Ann. Phys.}\ }\textbf {\bibinfo {volume}
  {323}},\ \bibinfo {pages} {2971} (\bibinfo {year} {2008})}\BibitemShut
  {NoStop}%
\bibitem [{\citenamefont {Braaten}\ \emph {et~al.}(2011)\citenamefont
  {Braaten}, \citenamefont {Kang},\ and\ \citenamefont
  {Platter}}]{Braaten2011Universal}%
  \BibitemOpen
  \bibfield  {author} {\bibinfo {author} {\bibfnamefont {E.}~\bibnamefont
  {Braaten}}, \bibinfo {author} {\bibfnamefont {D.}~\bibnamefont {Kang}}, \
  and\ \bibinfo {author} {\bibfnamefont {L.}~\bibnamefont {Platter}},\ }\href
  {\doibase 10.1103/PhysRevLett.106.153005} {\bibfield  {journal} {\bibinfo
  {journal} {Phys. Rev. Lett.}\ }\textbf {\bibinfo {volume} {106}},\ \bibinfo
  {pages} {153005} (\bibinfo {year} {2011})}\BibitemShut {NoStop}%
\end{thebibliography}%

\end{document}